\documentstyle[12pt,hpatex]{article}
\begin{document}
\title{ A New Kinematical Derivation of the Lorentz \newline Transformation
and the Particle Description of Light}  
\authors{J.H.Field}
\address{D\'{e}partement de Physique Nucl\'{e}aire et Corpusculaire
 Universit\'{e} de Gen\`{e}ve . 24, quai Ernest-Ansermet
 CH-1211 Gen\`{e}ve 4.}
 Published in Helv. Phys. Acta. {\bf 70} (1997) 542-564 

\abstract{ 
 The Lorentz Transformation is derived from only three simple postulates: (i) a weak
 kinematical form of the Special Relativity Principle that requires the equivalence
 of reciprocal space-time measurements by two different inertial observers; (ii)
 Uniqueness, that is the condition that the Lorentz Transformation should be a single valued function of 
 its arguments; (iii) Spatial Isotropy.
 It is also shown that to derive the Lorentz Transformation for space-time points lying along a
  common axis, parallel to the relative velocity direction, of two 
 inertial frames, postulates (i) and (ii) are sufficient.
  The kinematics of the Lorentz Transformation is then developed
 to demonstrate that, for consistency with Classical Electrodynamics,
  light must consist of massless (or almost massless) particles: photons.
}
\section{Introduction}
 In his seminal paper of 1905 on Special Relativity ~\cite{x1} Einstein derived
 the Lorentz Transformation from two main ~\cite{x2} postulates:
 \begin{itemize}
 \item[1.] The laws of physics are the same in all inertial frames (the Special Relativity
 Principle).
 \item[2.] In any given inertial frame, the velocity of light is a constant, c, 
 independant of the velocity of the source (Einstein's second postulate). 
 \end{itemize}
 The first postulate was stated by Galileo ~\cite{x3} and was well known, before
 the advent of Special Relativity, to be respected by the laws of Classical
 Mechanics. It is clear that Einstein regarded the second postulate as a 
 `law of physics' and so, in fact, a special case of the first postulate. 
 Whether this is {\it necessarily} the case will be discussed below in Section 3
  \par A study of Einstein's derivation of the Lorentz Transformation shows that the only `law of 
  physics' that is involved is that which is implicit in the very definition
  of an inertial frame: Newton's First Law. It will be demonstrated from purely
  kinematical arguments in Section 3. below that Einstein's `light signals'
  may be identified with particles moving in straight lines with fixed momentum 
  and energy according to Newton's First Law. That no other laws of physics are
  involved is crucial for the significance of Einstein's derivation of the
  Lorentz Transformation, and for the meaning of Special Relativity. What separates clearly Einstein's
  achievement from the related work of Lorentz ~\cite{x4}, Larmor ~\cite{x5}
  and Poincar\'{e} ~\cite{x6} is his realisation that the Lorentz Transformation gives the relation 
  between the space-time geometries (or, in momentum space, the kinematics)
  of different inertial frames. These are aspects of physics which underlie,
  but are quite distinct from, the actual dynamical laws. It turns out
  that these laws are indeed covariant under the Lorentz Transformation ~\cite{x7} and so respect 
  the Special Relativity Principle as 
  stated in Einstein's first postulate. However, as will be shown below, a much weaker statement
  of the Relativity Principle than Einstein's first postulate is sufficient
  to derive the Lorentz Transformation. More concretely it may be stated that the Lorentz Transformation describes
  only how space-time points or energy-momentum 4-vectors appear to
  different inertial observers, while the dynamical laws of physics, for
  example Newton's Second Law, the Lorentz Equation and Maxwell's Equations
  with sources, describe rather how future measurements of space-time points
  or, other 4-vectors, may be predicted from a knowledge of past or present
  ones ~\cite{x8}. In the works of Lorentz, Larmor and Poincar\'{e} on 
  electrodynamics, these two different aspects, the one kinematical the other
  dynamical, are inextricably interwoven.
  \par Because of this clear cut distinction in Special Relativity between
  kinematics and dynamics, it was recognised at an early date by
  Ignatowsky and Frank and 
  Rothe ~\cite{x9} that Einstein's second postulate was not necessary to
  derive the Lorentz Transformation. The questions then arise: what are the {\it weakest} postulates
  which are sufficient to derive it and what is their {\it minimum} number?
  The first part of this paper attempts to give an answer to these questions.
  As in other work on the subject, a purely kinematical approach is adopted
  without any reference, in the derivation, to Classical Electrodynamics or
  any other dynamical law of physics. The kinematical consequences of the Lorentz Transformation
  are then compared to results of Classical Electrodynamics to establish the
  identity of the velocity parameter that necessarily appears in the Lorentz Transformation and
  the velocity of light.
  A consistent interpretation then requires light to consist of
  massless (or almost massless) particles ~\cite{x10}. 
  \par The plan of the paper is as follows: In the following Section the three 
  postulates on which the derivation of the Lorentz Transformation is based are introduced. Then
  separate derivations of the Lorentz Transformation and the Parallel Velocity Addition Relation
  (PVAR) are given. In Section 3 the kinematical consequences of the Lorentz Transformation are 
  developed and it is shown that any physical object whose mass equivalent
  is much less than its energy will be observed to have a 
  constant velocity $V$ in any inertial frame.
  Special cases are photons ($V=c$) and massless or very light neutrinos.
  In the final Section the derivation of the Lorentz Transformation given here is
  compared with other similar work in the literature and conclusions are given.
  \section{Derivation of the Lorentz Transformation and the Parallel
 Velocity Addition Relation}
 The first postulate used (Postulate A) is a kinematical version of the Special Relativity
 Principle: 
 \begin{itemize}
 \item[A.] {\bf Reciprocal Space-Time Measurements (STM) of similar measuring rods and 
 clocks in two different inertial frames S and S', by observers at rest in these
 frames, give identical results.}     
 \end{itemize}
 The frame S may be identified with Einstein's `stationary system' ~\cite{x1},
 while, without loss of generality, S' may be assumed to move along the common
 x-axis of S and S' with velocity $v$. The y, z axes of S, S' are also taken
 to be parallel. Two examples of `reciprocal measurements' ~\cite{x11},~\cite{x12} 
 are given below. As discussed in Section 4, some recent derivations of the
  Lorentz Transformation
 based upon sophisticated gedankenexperimente have, implicitly or explicitly,
 used Postulate A. Unlike in Einstein's first postulate there is no mention in Postulate A of
 the `laws of physics'. Newton's First Law is however implicit in the term
 `inertial', which means that the frames `remain in the same state of uniform
 rectilinear motion' ~\cite{x13}. It may be objected that the `laws of physics'
 are implicit in the physical processes underlying the operation of the clocks.
 A mechanical clock relies on the dynamical laws of Classical Mechanics, an 
 atomic clock on those of Quantum Mechanics. By assuming spatial isotropy
 (see below) it can however be guaranteed that the clocks in, say, S and S', are
 identical even though the acceleration necessary to give S and S' their 
 relative motion may change the operation of the clocks in some unknown way.
 For example, suppose that the clocks of identical construction in S and S'
 are originally at rest in a third inertial frame \~{S} where they are 
 synchronised by any convenient procedure. If the relative velocity beween
 S and S' is now produced by giving the frames S, S' (and their associated 
 clocks) equal and opposite uniform accelerations for a suitably chosen time in
 \~{S}, it is clear, from the spatial symmetry required by the spatial
 isotropy postulate, that the clocks, originally identical in \~{S}, must
 remain so in S and S'.
 \par The second postulate (Postulate B) is that of Uniqueness. This has been previously
 used ~\cite{x14} in the derivation of the PVAR. To the best of the writer's
 knowledge, it is here applied for the first time in the derivation of the Lorentz Transformation
 itself. This postulate is based on the hypothesis that if an observer in S
 performs a number N of STM then another, similarly equipped, observer at rest 
 in S' can also make N STM in one-to-one correspondence with those made by the
 observer in S ~\cite{x15}. This will be so provided that the Lorentz Transformation is a single-valued function
 of its arguments. In the contrary case that the Lorentz Transformation is not single-valued then
 one STM measurement in S may correspond to several in S' of vice-versa. Such
 an asymmetrical situation between two inertial frames is clearly at variance
 with the Principle of Relativity. The Uniqueness postulate will also be used 
 to derive, separately, the PVAR, without assuming, as was done in Ref.[14],
 Einstein's second postulate. The statement of the Uniqueness postulate is:
\begin{itemize}
\item[B.] {\bf If $f(\chi,\xi,\zeta,..)=0$ represents either a Lorentz Transformation equation or the PVAR,
then $\chi$ must be a real single-valued function of $\xi$, $\zeta$,... ; $\xi$ must
be a real single-valued function of $\chi$, $\zeta$,... and so on for each of the 
arguments of $f$.} 
\end{itemize}         
For the PVAR, which has just 3 arguments, a sufficient ~\cite{x16} condition is that
$f$ should be trilinear in the relative velocities $v_{AB}$, $v_{BC}$, $v_{CA}$ of three
inertial frames A, B, C:
\begin{eqnarray}
0 & = & P + Q_1v_{AB}+Q_2v_{BC}+Q_3v_{CA}  \nonumber \\
  &   & +R_1v_{AB}v_{BC}+R_2v_{AB}v_{CA}+R_3v_{BC}v_{CA}+Sv_{AB}v_{BC}v_{CA}
\end{eqnarray}      
The coefficients $P$, $Q_i$, $R_j$, S are constants. If any two of $v_{AB}$, $v_{BC}$,
$v_{CD}$ are fixed then Eqn.(2.1) is linear in the remaining variable, and so has a unique 
solution ~\cite{x17}.
\par The third postulate (Postulate C), spatial isotropy, requires no special comment:
\begin{itemize}
\item[C.] {\bf The Lorentz Transformation equations must be independent of the directions of the spatial axes used
to specify a STM.}
\end{itemize} 

\par These three postulates are now used to derive the Lorentz Transformation. In a first step it is
assumed that the STM lie on the common x-axis of the frames S and S'. The generalisation
to $y \neq 0$, $z \neq 0$ will be made subsequently. When $y=y'=z=z'=0$ the Lorentz Transformation has the form:
\begin{equation}
x'=f(x,t)
\end{equation}
Postulate B is satisfied provided that Eqn.(2.2) can be written as:
\begin{equation}
x'+a_1x+a_2t+b_1xx'+b_2xt+b_3x't+cxx't=0
\end{equation}
where $a_i$, $b_j$, $c$ are independent of $x$, $x'$, $t$. The velocity of S' relative to S is:
\begin{equation}
 v \equiv \left. \frac{dx}{dt} \right|_{x'=\chi'}
\end{equation} 
where $\chi'$ may take any constant value. Differentiating Eqn.(2.3) with respect to t and 
using Eqn.(2.4) gives:
\begin{equation}
 v = - \frac{a_2+b_2x+b_3\chi'+c\chi'x}{a_1+b_1\chi'+b_2t+c\chi't}
\end{equation} 
Since Eqn.(2.5) must hold for all values of $x$, $\chi'$, $t$ it follows that:
\begin{equation}
 b_1=b_2=b_3=c=0
\end{equation}
so that:
\begin{equation}
 v = - \frac{a_2}{a_1}
\end{equation} 
Using Eqns.(2.6),(2.7) Eqn(2.3) may be written as:
\begin{equation}
 x'= \gamma (x-vt)
\end{equation} 
where
\begin{equation}
 \gamma \equiv -a_1
\end{equation} 
\par The Lorentz Transformation inverse to Eqn(2.2) is of the form:
\begin{equation}
x=f'(x',t')
\end{equation}
The velocity of S relative to S', $v'$, is defined as:
\begin{equation}
 v' \equiv -\left. \frac{dx'}{dt'} \right|_{x=\chi}
\end{equation} 
where $\chi$ may take any constant value. In many derivations of the Lorentz Transformation (including 
Einstein's in Ref.[1] ) it is assumed that:
\begin{equation}
 v'= v
\end{equation}
This hypothesis is called the Reciprocity Postulate. It has been proved by Berzi and Gorini
~\cite{x18} to be a consequence of the Special Relativity Principle and the usual postulates
of space-time homogeneity and spatial isotropy. Since, in the present derivation, space-time
homogeneity is not assumed, Eqn.(2.12) cannot be assumed to be correct. It will now be shown 
however, that the Reciprocity Postulate is a necessary consequence of Postulate A and Postulate B alone. That is,
it is independant of the assumed properties of space-time in the case that Postulate B is true.
Repeating the line of argument leading from Eqn.(2.2) to Eqn.(2.8), but starting instead
with Eqn.(2.10), gives:
\begin{equation}
 x= \gamma' (x'+v't')
\end{equation} 
Suppose now that a measuring rod of unit length, lying along the Ox' axis, and at rest in S'
is observed, at fixed S time $t$, by S. It follows from Eqn.(2.8) that S will observe 
the length of the rod to be $l$ where:
\begin{equation}
 l= \frac{1}{\gamma}
\end{equation}
If S' now makes a reciprocal measurement of a similar rod at rest in S the length will
 be observed to be $l'$, where, from Eqn.(2.13):
\begin{equation}
 l'= \frac{1}{\gamma'}
\end{equation} 
Using Postulate A:
\begin{equation}
 l = l'
\end{equation}
so that
\begin{equation}
 \gamma = \gamma'
\end{equation}
Hence, using Eqns.(2.8), (2.13), (2.17) the Lorentz Transformation may be written as:
\begin{eqnarray}
x' & = & \gamma (x-vt)    \\
t' & = & \frac{\gamma v}{v'}(t-\frac{x \delta}{v})
\end{eqnarray}
where
\begin{equation} 
\delta \equiv \frac{(\gamma^2-1)}{\gamma^2}
\end{equation}
The inverse transformations are:
\begin{eqnarray}
x & = & \gamma (x'+v't')    \\
t & = & \frac{\gamma v'}{v}(t'+\frac{x' \delta}{v'})
\end{eqnarray}
Consider now a clock at rest in S' located at $x'=0$. As seen from S, the position of
the clock, after the time $t=\tau_0$, is given by:
\begin{equation} 
x = v\tau_o
\end{equation}
From Eqns.(2.19) and (2.23) the elapsed time $\tau'$ indicated by the clock, as 
observed from S, during the interval $\tau_0$ of S frame time, is:
\begin{equation} 
\tau' = \frac{\gamma v}{v'}\tau_0(1-\delta)
\end{equation}
If an observer in S' makes now a reciprocal observation of a similar clock at rest in S,
then the position of the clock, after the time $t'=\tau_0$, is given by:
\begin{equation} 
x' = -v'\tau_o
\end{equation}
Combining Eqns.(2.22) and (2.25): 
\begin{equation} 
\tau = \frac{\gamma v'}{v}\tau_0(1-\delta)
\end{equation}
where $\tau$ is the elapsed time, indicated by the clock at rest in S, as observed from S'
during the period $\tau_0$ of S' time. Because the two measurements of the elapsed time
indicated by the similar clocks are reciprocal:
\begin{equation} 
 \tau = \tau'
\end{equation}
It then follows from Eqns.(2.24) and (2.26) that:
\[ v = v'\]
The alternative solution with $v'=-v$ may be rejected since it corresponds to
the case where the clock in S' runs backwards in time ( $t' \rightarrow -t'$ ).
In the case of two inertial frames equipped with identical clocks $\Delta t$ must 
 have the same sign in both frames. Thus the Reciprocity Postulate Eqn.(2.12) is
 a consequence of Postulate A and Postulate B only. Using now Eqn.(2.12) the Lorentz Transformation in Eqns.(2.18) and (2.19)
  becomes:
 \begin{eqnarray}
x' & = & \gamma (x-vt)    \\
t' & = & \gamma (t-\frac{x \delta}{v})
\end{eqnarray}
Since the Lorentz Transformation is completely defined by the relative velocity $v$
 between the frames
S and S' it follows that the unknown parameter $\gamma$, (and hence, from Eqn.(2.20)
 $\delta$ ) must be a function of $v$.  
\par Suppose now that a physical object moves with velocity $u$ in the direction
of the positive x' axis in S'. Its velocity $w$ in the direction of the positive
 x axis, as observed in S, can be derived by differentiating Eqns.(2.28),(2.29)
 with respect to $t$ and using the definitions: 
\begin{eqnarray}
w & \equiv & \frac{dx}{dt}    \\
u & \equiv & \frac{dx'}{dt'} 
\end{eqnarray} 
The ratio of Eqn.(2.28) to Eqn.(2.29) after differentiation, use of Eqns.(2.30),
 (2.31) and some rearrangement, gives:
\begin{equation} 
 w = \frac{u+v}{1+\frac{u \delta(v)}{v}}
\end{equation} 
This is the Parallel Velocity Addition Relation (PVAR). By making use of the 
Reciprocity Postulate and Postulate C (Spatial Isotropy) it has been demonstrated 
~\cite{x19} that the PVAR is symmetric under the exchange $ u \leftrightarrow v$.
As will be now shown, 
this symmetry is in fact a consequence of the Reciprocity Postulate alone.
 Introducing the notation:
 \[ v_{AB}=v~~~v_{BC}=u~~~v_{AC}=w \] 
 Eqn.(2.32) gives:
 \begin{equation} 
 v_{AC} = \frac{v_{BC}+v_{AB}}{1+\frac{v_{BC} \delta(v_{AB})}{v_{AB}}}
\end{equation} 
Exchanging the labels A and C in Eqn.(2.33) :
 \begin{equation} 
 v_{CA} = \frac{v_{BA}+v_{CB}}{1+\frac{v_{BA} \delta(v_{CB})}{v_{CB}}}
\end{equation}
Using the Reciprocity Postulate, $ v_{CA} =  - v_{AC} $ etc, Eqn.(2.34) may be written
as :
 \begin{equation} 
 v_{AC} = \frac{v_{AB}+v_{BC}}{1+\frac{v_{AB} \delta(-v_{BC})}{v_{BC}}}
\end{equation}
Comparing Eqns. (2.33) and (3.35) gives the relation:
 \begin{equation} 
 \frac{v^2}{\delta(v)} = \frac{u^2}{\delta(-u)} = \pm V^2 
\end{equation} 
Since $v$, $u$ are independant variables the first two members of Eqn.(2.36)
must each be equal to the universal constant $\pm V^2$. By inspection, $V$ has the 
dimensions of velocity. Setting $u = -u$ in Eqn.(2.36) it follows that :
 \begin{equation} 
 \delta(-u) = \delta(u)
\end{equation} 
and hence, from Eqn.(2.36) that:
 \begin{equation} 
 \frac{u}{v}\delta(v) = \frac{v}{u}\delta(u)
\end{equation} 
 The PVAR, Eqn.(2.32), is thus a symmetric function of $u$ and $v$.
 Two distinct possiblities now exist
 on combining Eqns.(2.20), (2.32) and (2.36) ~\cite{x21}:
\begin{itemize}
\item[a)] plus sign in Eqn.(2.36):
\begin{eqnarray}
\gamma(v) &  = & \frac{1}{\sqrt{1-(\frac{v}{V})^2}}     \\
w & = & \frac{u+v}{1+\frac{uv}{V^2}} 
\end{eqnarray}    
\item[b)] minus sign in Eqn.(2.36):
\begin{eqnarray}
\gamma(v) &  = & \frac{1}{\sqrt{1+(\frac{v}{V})^2}}     \\
w & = & \frac{u+v}{1-\frac{uv}{V^2}} 
\end{eqnarray} 
\end{itemize} 
Case b) gives no restriction on the possible values of $u$ and $v$. However,
in this case, the PVAR Eqn.(2.42) is not a single valued function of its
arguments. It is thus excluded by Postulate B (Uniqueness). To show this, it may be noted
that, for any value of $v$ , a value of $u$, $u_{\infty}$ may be chosen such that
$w$ is infinite:
\begin{equation} 
 u_{\infty} =\frac{V^2}{v}
\end{equation} 
defining
\begin{equation} 
 \Delta \equiv u_{\infty}-u 
\end{equation} 
Eqn.(2.42) may be written as:
\begin{equation} 
 w = -\frac{(v+u_{\infty}+\Delta) u_{\infty}}{ \Delta}
\end{equation} 
 Since $w \rightarrow -\infty$ as $\Delta \rightarrow 0$ when $\Delta > 0$,
 whereas $w \rightarrow +\infty$ as $ \Delta \rightarrow 0$ when $\Delta < 0$,  
 Eqn.(2.45) does not give a unique solution for $w$ when $u = u_{\infty}$.
 \par In case a) the requirement that $\gamma$ should be a real quantity (Postulate B)
 shows that, in this case, $V$ plays the r\^{o}le of a limiting velocity:
 \begin{equation} 
 u^2,v^2 \leq V^2 
\end{equation}
With the definition:
 \begin{equation} 
 u_{lim} \equiv -\frac{V^2}{v}  
\end{equation}
 the restrictions on $u$, $v$ given by Eqn.(2.46) imply 
either that no value, or a unique value, of  $u_{lim}$ exists.
A consequence of Eqn.(2.47) is:
\begin{equation} 
 \frac{|u_{lim}|}{V} = \frac{V}{|v|}    
\end{equation}    
If $|v| < V$, then $|u_{lim}| > V$, in contradiction with Eqn.(2.46). For
the case $v = V = - u_{lim}$ Eqn.(2.40) gives the result: 
$w = - u_{lim} = V$. Hence, for all values of $u$, $v$ consistent with
Eqn.(2.46), the PVAR Eqn.(2.40) gives a unique value of $w$, and so verifies Postulate B.
Thus when $v=V$, $w$ takes also the value $V$, independently of the value of $u$.
 From the symmetry of Eqn.(40) this statement remains true when $u$ and $v$ are
 interchanged. A consequence is that if a physical object has velocity $V$ in any
 inertial frame (say S', when $u=V$) then it has the velocity $V$ in any other
 inertial frame (say S , when $w=V$). The physical interpretation of $V$ is then the limiting velocity 
 (independent of the choice of inertial frame) which any physical object
 may attain. It can already be seen that V has the same property as that ascribed
 to the velocity of light $c$ in Einstein's second postulate. In Section 3. below the
 limiting velocity
 $V$ will be related to the mass, energy and momentum of any physical object.
\par The Lorentz Transformation for STM lying along the common x axis of S, S' has now been completely
determined by postulates A and B only. It corresponds to case a) above
(plus sign in Eqn.(2.36)) and is given by the equations:
 \begin{eqnarray}
x' & = & \gamma (x-vt)    \\
t' & = & \gamma (t-\frac{vx}{V^2}) \\
\gamma & = & \frac{1}{\sqrt{1-(\frac{v}{V})^2}}
\end{eqnarray}  
\par This result is now generalised to include STM lying outside the common
x-axis of S and S'.
In this case Postulate C is also required. 
 Considering first STM with $y \neq 0, z=0$, then Postulate B implies
that Eqn.(2.49) should be modified to:
\begin{equation}
x'  =  \gamma (x-vt)+yf(x',x,t)
\end{equation}
where the function $f$ is trilinear in $x'$, $x$, $t$. Postulate C requires that Eqn.(2.52)
should be invariant under the operation $y \rightarrow -y$ giving:
\begin{equation}
x'  =  \gamma (x-vt)-yf(x',x,t)
\end{equation}
Subtracting Eqn.(2.53) from Eqn.(2.52):
\begin{equation}
 yf(x',x,t) = 0 
\end{equation}
Since this equation must hold for all values of $x'$, $x$, $y$, $t$ it
follows that: 
\[  f(x',x,t) = 0 \]
Thus Eqn.(2.49) is valid also for STM with $y \neq 0$. Using Postulate C it must also hold
 for STM with  $z \neq 0$.
\par Consider now the transformation equations for STM along the y-axis
($z=z'=0$) at $x' = 0$:
\[  f(y',y,x,t) = 0 \]
Postulate B is verified provided that $f$ is quadrilinear in $y'$, $y$, $x$, $t$ i.e.
\begin{eqnarray}
y' & + & A_1y+A_2x+A_3t+B_1y'y+B_2y'x+B_3y't \nonumber   \\
   & + & B_4yx+B_5yt+B_6xt+C_1y'yx+C_2y'yt \nonumber \\
   & + & C_3y'xt+C_4yxt+Dy'yxt = 0
\end{eqnarray}  
Invariance under $x \rightarrow -x$ (Postulate C) implies that the coefficients of all 
terms containing $x$ must vanish. Similarly, invariance under
the combined transformation: $y \rightarrow -y$,
 $y' \rightarrow -y'$, gives the further conditions:
 \[ A_3 = B_1 = C_2 = 0 \]
 so that Eqn.(2.55) reduces to
\begin{equation}
 y' +A_1y+B_3y't+B_5yt = 0  
\end{equation}  
Fixing $y'$ to be equal to $\xi'$ and differentiating Eqn.(2.56) with respect to $t$
gives:
\begin{equation}
 A_1\left.\frac{dy}{dt}\right|_{y'=\xi'}+B_3\xi'+B_5(\left.\frac{dy}{dt}\right|
 _{y'=\xi'} +y) = 0  
\end{equation}
But, because the velocity of S' is perpendicular to y:
\[ \left.\frac{dy}{dt}\right|_{y'=\xi'}= 0 \]
so that Eqn.(2.57) becomes:
\[ B_3 \xi' + B_5y = 0 \]
As this equation must be true for arbitary $\xi'$, $y$ then: 
\[ B_3 = B_5 = 0 \]
giving, with Eqn.(2.56):
\begin{equation}
 y'= -A_1y  
\end{equation}
The proof that $A_1 = -1$ was given in Ref.[1]. Denote by $A_1(v)$ the coefficient in
Eqn.(2.58) corresponding to the Lorentz Transformation of Eqn.(2.49). The corresponding coefficient for the
Lorentz Transformation inverse to Eqn.(2.49) is then $A_1(-v)$. Applying, in succession, the Lorentz Transformation of
 Eqn.(2.49) and its inverse then Eqn.(2.58) gives:
\begin{equation}
  A_1(-v)A_1(v) = 1 
\end{equation} 
As the direction of the relative velocity of S and S' is perpendicular to the y, y'
axes $A_1(v)$ cannot depend on the spatial orientation of the relative velocity (Postulate C).
Hence:
\begin{equation}
  A_1(v)=A_1(-v)  
\end{equation}
Eqns.(2.59) and (2.60) give
\begin{equation}
  A_1(v)= \pm 1  
\end{equation}
Since, evidently, Eqn(2.57) reduces to $y = y'$ in the limit $v =0$, the minus sign 
must be taken in Eqn.(2.61) so that, for arbitary $v$, Eqn.(2.58) becomes:
\[ y' = y\] 
Application of Postulate C then implies that:
\[ z' = z\] 
The final result for the Lorentz Transformation of STM at an arbitary spatial position in S is then:
 \begin{eqnarray}
x' & = & \gamma (x-vt)    \\
y' & = & y   \\
z' & = & z   \\  
t' & = & \gamma (t-\frac{vx}{V^2})
\end{eqnarray}   
where $\gamma$ is defined in Eqn.(2.51).
The derivation has used Postulates A, B and C. However, as mentioned above, for the case
$ y=y'=z=z'=0$ Eqns.(2.62),(2.65) may be derived from Postulates A and B only.
\par Finally, in this Section, an alternative derivation of the PVAR is given using
only Postulate B, and the Reciprocity Postulate. As already noted, Postulate B is verified if the 
PVAR has the trilinear form of Eqn.(2.1).The argument given above to show that
the PVAR is symmetric under the exchange $u \leftrightarrow v$ is easily extended
to prove that it is symmetric under the exchange of any two of $u, v, w$
(see also Ref.[19]).
  This has the consequence that, in Eqn.(2.1):
\begin{equation}
 Q_1 = Q_2 = Q_3 \equiv Q   
\end{equation}
\begin{equation} 
 R_1 = R_2 = R_2 \equiv R   
\end{equation}
Imposing now the condition that $v_{CA}=0$ when $v_{AB}=v_{CB}$ (the Reciprocity
Postulate) gives:
\begin{equation}
 P-v_{AB}^2R = 0  
\end{equation}    
Since Eqn.(2.68) must hold for all values of $v_{AB}$ then
\begin{equation}
 P = R = 0  
\end{equation}
Using Eqns.(2.66), (2.67), (2.69) the PVAR may be written:
\begin{equation}
v_{AB}+v_{BC}+v_{CA}+\frac{S}{Q}v_{AB}v_{BC}v_{CA} = 0  
\end{equation}
From dimensional analysis of Eqn.(2.70) it follows that:
\begin{equation}
\frac{Q}{S} = \pm V^2  
\end{equation}
where $V$ is a universal constant with the dimensions of velocity.
From the definitions of $u$, $v$, $w$ given above, and using again the Reciprocity
Postulate, Eqn.(2.70) becomes identical to Eqn.(2.40) or (2.42) according as the
plus or minus sign respectively is chosen in Eqn.(2.71). The argument given above,
using the Uniqueness postulate (Postulate B) then eliminates the solution with the minus sign. 
Finally it may be remarked that if the Reciprocity Postulate is regarded, as is often
the case in the literature, as `obvious' it would follow that the PVAR has been derived
here purely from Postulate B, i.e. without recourse to the Special Relativity Principle
itself. In fact, as shown here and in Ref.[18], the Reciprocity Postulate is 
actually a necessary consequence of the Relativity Principle and other postulates
(space-time homogeneity and spatial isotropy in Ref.[18],
the Relativity Principle and Uniqueness in the present paper).
     \section{Kinematical Consequences of the Lorentz Transformation}
Introducing the notation $s \equiv Vt$, the Lorentz Transformation in Eqns.(2.62) to (2.65) may be written
 in the form:
 \begin{eqnarray}
x' & = & \gamma (x-\beta s)    \\
y' & = & y \\
z' & = & z \\ 
s' & = & \gamma (s-\beta x )
\end{eqnarray}        
where
\[ \beta \equiv \frac{v}{V}~~~~\gamma =\frac{1}{\sqrt{1-\beta^2}}  \]  
 The four component quantity $X \equiv (s ; x , y , z)$ is a 4-vector ~\cite{x22} whose 
 `length' $r_X$ is defined by the relation:
\begin{equation}
r_X^2 \equiv V^2\tau ^2 \equiv s^2-x^2-y^2-z^2
\end{equation}
As may be shown directly, using Eqns.(3.1)-(3.4), Eqn.(3.5) is invariant 
under the Lorentz Transformation S $\rightarrow$  S' corresponding to the replacement~\cite{x23}:
 \[ (s;x,y,z)~\rightarrow~(s';x',y',z') \]
The quantities $r_X$, $\tau$, with dimensions of length and time respectively, defined in
Eqn.(3.5) thus have the same value in all inertial frames, i.e. they are Lorentz invariant.
\par Consider now a physical object of Newtonian inertial mass (referred to subsequently
simply as `mass') situated at the space-time point X in S and moving with an arbitary
uniform velocity $\vec{u}$ in that frame. Suppose that the object is at rest at the origin
of the inertial frame S''. Using Eqn.(3.5) in the frame S'', it follows that if $t''$ is the
time at the object as observed in S'' then:
\begin{equation}
 \tau = t''
\end{equation}
That is the Lorentz invariant time $\tau$ is the {\it proper time} (the time in its own rest
frame) of the physical object. A new 4-vector $p$ may now be defined as:
\begin{equation}
 p \equiv m \frac{dX}{d \tau}
\end{equation}
 Because $\tau$ is Lorentz invariant $p$ transforms in the same way under the Lorentz Transformation
  as X~\cite{x24}. Indeed
 this is the property which defines, in general, a 4-vector. 
 Thus $p'$ observed in S' is related to $p$ observed in S by :
\begin{eqnarray}
p_x' & = & \gamma (p_x-\beta p_s)    \\
p_y' & = & p_y \\
p_z' & = & p_z \\ 
p_s' & = & \gamma (p_s-\beta p_x )
\end{eqnarray}
By considering the Lorentz Transformation parallel to $\vec{u}$, the infinitesimal time increments $\delta t$
 and $\delta \tau$ are related by the expression:
\begin{equation}
 \delta t = \gamma_u \delta \tau 
\end{equation}
where
\[ \beta_u \equiv \frac{u}{V}~~~~\gamma_u =\frac{1}{\sqrt{1-\beta^2_u}}  \]    
 Using Eqns.(3.7), (3.12) and taking the limits as $\delta t$, $\delta \tau$, 
 $\delta \vec{x}$, $\rightarrow 0$:
\begin{eqnarray}
p_x & = & m \frac{dx}{d \tau} = m \gamma_u \frac{dx}{dt} = m \gamma_u u_x    \\
p_y & = & m \gamma_u u_y  \\
p_z & = & m \gamma_u u_z  \\
p_s & = & m V \frac{dt}{d \tau} = m \gamma_u V 
\end{eqnarray}
Analagously to $r_X$, the length of $p$, $r_p$, is defined as: 
\begin{equation}
r_p^2 \equiv p_s^2-p_x^2-p_y^2-p_z^2 = m^2 V^2 \gamma_u^2 (1-\beta_u^2) = m^2 V^2
\end{equation}
 The mass $m$ is then
proportional to the length $r_p$ of the energy momentum 4-vector $p$
and $U \equiv ( \gamma_u V; \gamma_u u_x, \gamma_u u_y,  \gamma_u u_z )$
is also a 4-vector, the relativistic generalisation of the velocity vector of classical 
mechanics. Introducing the definitions:
\begin{eqnarray}
  p^2 & \equiv & p_x^2+p_y^2+p_z^2  \\
  E   & \equiv & V p_s
\end{eqnarray}
then Eqns.(3.13)-(3.17) lead to the relations
\begin{eqnarray}
\beta_u & = & \frac{Vp}{E} \\
\gamma_u & = & \frac{E}{m V^2} \\ 
  E^2  &  = &  p^2 V^2 + m^2 V^4 
\end{eqnarray}
To see the connection between $p$, $E$ and the physical quantities of Classical Mechanics,
consider the limit where $u \ll V$ i.e. $ \beta_u \ll 1$:
\begin{eqnarray}
 p =m \gamma_u u & = & mu(1+\frac{1}{2} \beta_u^2+... )  \\
 & \simeq & mu = p^{(N)} \\ 
 E=(p^2 V^2+ m^2 V^4)^\frac{1}{2} &  = &  m V^2 (1+\frac{p^2}{2 m^2 V^2} +... ) \\
 & \simeq & m V^2+\frac{p^2}{2m} = m V^2 + T^{(N)}  
\end{eqnarray} 
Here $p^{(N)}$, $T^{(N)}$ denote the Newtonian momentum and kinetic energy respectively.
The success of Newtonian mechanics in the everyday world then indicates that V must be
very large as compared to the typical velocities encountered on the surface of the 
earth. The physical meaning of the relativistic energy E is given by setting p = 0 in
Eqn.(3.22): 
\begin{equation}
E_0 = E(p=0) = m V^2
\end{equation}
This equation states the equivalence of mass and energy. {\it The mass m of a physical
object is equivalent to its relativistic energy in its own rest frame $E_0$ }. 
\par As a consequence of the conservation of relativistic momentum
$\vec{p} = (p_x,p_y,p_z)$ the relativistic energy E, defined in Eqn.(3.19)
is also a conserved quantity. Suppose that an ensemble of $N$ physical 
objects with energy-momentum
4-vectors $p^{i,IN}$ , $(i=1,2,..N )$ interact in an inelastic way so
as to produce a different ensemble of $M$ physical objects with energy-momentum 4-vectors
$p^{j,OUT}$ , $(j=1,2,..M )$. Define 4-vectors $P$, $\tilde{P}$, $\Delta P$
as:
\begin{eqnarray}
P & \equiv & \sum_{i=1}^N p^{i,IN} \\
\tilde{P} & \equiv & \sum_{j=1}^M p^{j,OUT} \\
\Delta P & \equiv & \tilde{P}- P
\end{eqnarray} 
Momentum conservation gives:
\begin{equation}
\Delta P_x = \Delta P_y = \Delta P_z = 0
\end{equation}
Applying the Lorentz Transformation Eqn.(3.8) to $\Delta P_x$ :
\begin{equation}
\Delta P_{x'} = \gamma (\Delta P_x-\beta \Delta P_s)
\end{equation}
In order that $\Delta P_{x'}$ is zero for arbitary $\gamma$, $\beta$ 
(i.e. that momentum is always conserved in the frame S') it follows that
$\Delta P_s = 0$. Use of Eqns.(3.28-3.30) and (3.19) gives:
\begin{equation}
\sum_{i=1}^{N} E^{i,IN} = \sum_{j=1}^{M} E^{j,OUT}
\end{equation}
so that conservation of relativistic energy, E, is a consequence of the 
conservation of relativistic momentum.
\par Another, related, conserved quantity is the Lorentz invariant common
`effective mass' of the $N$ incoming or $M$ outgoing physical objects, 
defined by the relation:
\begin{equation}
M_{eff}^2V^2 \equiv P_s^2-P_x^2-P_y^2 -P_z^2 =
\tilde{P}_s^2-\tilde{P}_x^2-\tilde{P}_y^2 -\tilde{P}_z^2
\end{equation} 
In a way analagous to Eqn(3.27) for a single physical object, $M_{eff}$ is
equivalent to the total energy in the overall centre-of-mass system, where:
\[ \sum_{i=1}^{N} P^{i,IN} = \sum_{j=1}^{M} P^{j,OUT} = 0  \]
so that:
\begin{equation}
\sum_{i=1}^{N} E^{i,IN}_0 = \sum_{j=1}^{M} E^{j,OUT}_0 = M_{eff} V^2
\end{equation}
\par By using the Lorentz Transformation for energy-momentum 4-vectors Eqns.(3.8)-(3.11), and the
conservation laws of relativistic energy and momentum, it is not difficult to
devise simple particle physics experiments to determine the velocity parameter 
$V$. Two such examples follow:
\begin{itemize} 
\item[(i)] A proton with measured velocity $v_{IN}$ collides with a proton at 
rest. The recoil and scattered particles are required to have momentum 
vectors making equal angles $\theta$ with the direction of the incoming proton.
$V$ is then given by the relation:
\begin{equation}
V=\frac{v_{IN}[\cos^2 \theta+\cos 2 \theta]}{2 \cos \theta \sqrt{\cos 2 \theta}}
\end{equation}
For example, with $v_{IN} = 0.5 V$ , (proton momentum of 541 MeV/c, assuming
that $V = c$) then $\theta = 43.93^{\circ}$.
\item[(ii)] An annihilation photon from a para-positronium atom at rest 
Compton scatters on a free electron. The recoil electron and the
scattered photon have momentum vectors making equal angles with the incoming
photon direction. If $\beta_{OUT}= v_{OUT}/V$ where $v_{OUT}$ is the velocity
of the recoil electron, then:
\begin{equation}
\frac{1+\beta_{OUT}}{\sqrt{1-\beta_{OUT}^2}} = 2
\end{equation}
Solving Eqn.(3.37), it is found that:
\begin{equation}
V = 1.412 v_{OUT}
\end{equation}
\end{itemize} 
 \par For any physical object whose relativistic momentum is much larger than
 the equivalent of its mass:
\begin{equation}
 p \gg mV
\end{equation}
it follows from Eqns.(3.20),(3.22) that:
\begin{equation}
 u \simeq V
\end{equation}
\par all highly relativistic objects, in the sense of Eqn.(3.39) then have a
velocity slightly smaller than, but very close to, $V$. This has
 been demonstrated here for such objects in observed from the frame S. 
 It was shown in the previous Section however, by use of the PVAR, that the
 same is true in all inertial frames. A special case of Eqn.(3.40), corresponding
 to exact equality, is a massless particle, such as a photon or a neutrino, when:
\begin{equation}
 u = V = c
\end{equation}
If photons are massless particles then Einstein's second postulate has indeed been
shown to be a necessary consequence of Postulates A, B and C. However, as previously pointed
out ~\cite{x25}, the actual value of the photon mass is an experimentally determined parameter
 ~\cite{x26}, like the mass of any other particle. If it is admitted that the photon may have
 a non-zero mass, smaller than the experimental upper limit, then, for sufficiently low 
 energy photons, Einstein's second postulate would no longer hold, and so his derivation of
 the Lorentz transformation would not, in this case, be valid. No such restrictions apply to
 purely kinematical derivations, such as those originally reported in Ref.[9], or that presented
 in the present paper.
 \par It may be objected that the 4-vector definition in Eqn.(3.7) makes no sense if $m = 0$,
 as in this case the rest frame S'' of the particle cannot be defined. Actually, however the 
 quantity $\gamma_u m$ which occurs as a factor in all the components of $p$ (see Eqns.(3.13)
 -(3.16)) has a finite limit as $m \rightarrow 0$, $\gamma_u \rightarrow \infty$. Using 
 Eqns.(3.20),(3.22) one obtains:
\begin{eqnarray}
\gamma_u m & = & m [1-\beta_u^2]^{-\frac{1}{2}} \nonumber \\
  & = & m \left[ 1-\frac{V^2p^2}{V^2p^2+m^2 V^4 }\right]^{-\frac{1}{2}} \nonumber \\
  & = & [ \frac{p^2}{V^2}+m^2}]^{\frac{1}{2}
\end{eqnarray}        
The right hand side of Eqn.(3.42) is finite as $m \rightarrow 0$. It is crucial in discussing
the massless limit of Eqn.(3.42) that $m$ is identified with the Lorentz invariant 
Newtonian inertial mass. As emphasised by Okun ~\cite{x27} many texts and popular books on
Special Relativity still introduce a velocity dependent mass ~\cite{x28} that is, in fact
defined as the quantity on the left hand side of Eqn.(3.42). One then arrives at the somewhat
paradoxical conclusion that the relativistic mass of an object whose Newtonian mass is zero,
is not zero. In fact, for highly relativistic objects, the `relativistic mass' is simply
equal to $p/V$, an already well defined kinematical quantity, so that the use of the term
 `relativistic mass' becomes redundant. In all cases, (and, suprisingly, in contradiction 
 with what is claimed in Ref.[28]), the physical meaning of the equations of relativistic
 kinematics is much more transparent when only the Newtonian inertial mass is used in them
 ~\cite{x27}. 
 \par  A simple conclusion on the nature of
 light, and its relation to all other physical objects existing in nature can now drawn.
 If photons are indeed massless (~or effectively massless in the sense of Eqn.(3.39)~)
  objects moving with constant momentum according to Newton's
 First Law, then:
 \begin{itemize}
 \item[(1)] The `aether' becomes completely redundant. Since light is not (any more than any other
 physical object in motion) a wave phenomenon in the classical sense,
 then no wave-carrying medium is required. The
 Wave Theory of Light in the 19th Century, and the confusion over `wave particle duality' in the
 early part of the 20th, arose because the mathematical description of large ensembles of
 photons, each of which is {\it individually} described by the laws of Quantum Mechanics, 
 has a structure very similar to that of transverse waves in a classical medium.
 The mathematical descriptions are isomorphic even though the physical systems are quite 
 distinct. This point is further elaborated elsewhere ~\cite{x29}.
 \item[(2)] The counter-intuitive nature of the PVAR, which, as Einstein realised, (see for 
 example the popular book referred to in Ref.[12]) is a major stumbling block for the
 understanding of Special Relativity, is not due to some specific and mysterious property
 of light, as seems to be the case when it embodied in Einstein's second postulate, but
 a property of the geometry of space-time that relates observations of {\it all physical
 objects} with $ p \gg mV$. This fact may appear just as mysterious. It is, however, an
 inevitable logical consequence of simple, apparently self-evident, postulates such as 
 those used above or in other purely kinematical derivations of the Lorentz Transformation.
 \end{itemize} 
\section{Comparison with Previous Work and Conclusions}
The literature on the derivation of the Lorentz Transformation is vast (see Ref.[2] of Ref[18] for a partial
list of work published before 1968). Here only a brief survey of some more recent work is 
presented to put in perspective the derivation given in the present paper. Firstly, a list of 
the different postulates
that have been used in the literature is given. The postulates used in several different
derivations are then presented in Table I. For reference, Einstein's derivation of 1905 is 
included as the first entry. The other derivations in Table I do not make use of Einstein's
second postulate.
\par The postulates, and the abbreviation by which they are referred to in Table I are:
\begin{itemize} 
\item The Special Relativity Principle (SRP)
\item The Reciprocity Postulate (RP)
\item Space-Time Homogeneity (STH)
\item Spatial Isotropy (SI)
\item Relativistic Transverse Momentum Conservation (TMC)
\item The Group Property (GP)
\item Relativistic Mass Increase (RMI)
\item The Constancy of the Velocity of Light (CVL)
\item The Sign of the Limiting Velocity Squared (SLV2)
\item The Causality Postulate (CP) 
\item The Uniqueness Postulate (UP)
\end{itemize}
It can be seen from Table I that all authors cited require the Relativity Principle.
Similarly, all authors, prior to the present paper, require Space-Time Homogeneity. 
It has been shown ~\cite{x25,x30,x34} that this requires the Lorentz Transformation to be linear. Since
Einstein assumed linearity, without proof, Space-Time Homogeneity is included among
Einstein's necessary postulates in Table I. In the present paper it is shown that the
linearity of the Lorentz Transformation (Eqns.(2.8),(2.13)) follows from the Uniqueness postulate and the
definitions (Eqns.(2.4),(2.11) of relative velocity alone, i.e. no explicit postulates on
space-time geometry are required. All the derivations in Table I require, however, Spatial
Isotropy to derive the Lorentz Transformation for space time points lying outside the $x$, $x'$ axis.
Since the Reciprocity Postulate has been shown ~\cite{x18} to be a consequence of the Special
 Relativity Principle, Space-Time Homogeneity and Spatial Isotropy, it is not included in Table I 
 as these three postulates are included in all the derivations cited, except that of the present
 paper.
\begin{table}
\begin{center}
\begin{tabular}{||c||c||c|c|c|c|c|c|c|c|c|c||} \hline\hline
Author & Ref. & SRP &  STH & SI & TMC & GP & RMI & CVL & SLV2 & CP & UP \\ \hline\hline
Einstein & [1] & Y & Y & Y & N & N & N & Y & N & N & N \\ \hline
Terletskii & [30] & Y & Y & Y & N & Y & Y  & N & N & N & N \\ \hline
Weinstock & [31] & Y & Y & Y & Y & N & Y  & N & N & N & N \\ \hline
Lee and Kalotas & [32] & Y & Y & Y & N & Y & N  & N & Y & N & N \\ \hline
L\'{e}vy-Leblond & [25] & Y & Y & Y & N & Y & N  & N & N & Y & N \\ \hline
Sen  & [33] & Y & Y & Y & N & N & N  & N & Y & N & N \\ \hline
This paper  & - & Y & N & Y & N & N & N & N & N & N & Y \\ \hline\hline
\end{tabular}
\caption[]{ Postulates required for different derivations of the Lorentz Transformation.
 Y $\equiv$ `YES' (postulate used) , N $\equiv$ `NO' (postulate not used). See the text
 for the definitions of the postulates. All derivations except that of this paper and 
L\'{e}vy-Leblond also assumed the Reciprocity Postulate. Since however the latter has
been shown~\cite{x18} to be a consequence of SRP, STH and SI which are assumed in every
case in these derivations it is omitted from the list of {\it required} postulates in
the  Table.  }
\end{center}
\end{table} 
\par  The postulates of Constancy of the Velocity of Light, and Conservation of Relativistic
Momentum used in Refs.[1,31] respectively are `strong' consequences of the Lorentz Transformation that must be
guessed or derived directly from experiment. If the Group Property~\cite{x35} of the
transformation is used, as in Refs.[9,25,30,32] then such strong postulates are not
necessary. The present paper gives however a counter example to show that the Lorentz Transformation
may be derived, in the absence of strong postulates, without assuming the Group
Property. The remaining postulates cited in Table I ( Relativistic Mass Increase, Sign of the
Limiting Velocity Squared, and the Causality Postulate) are introduced to circumvent a
problem which arises in all derivations that do not use Einstein's second postulate.
For dimensional reasons, a constant with the dimensions of velocity squared must occur
in the equations (Eqn.(2.36) in the present paper), but its sign is not specified.
One choice leads to the Lorentz Transformation, when the constant is identified with $c^2$, the other to
another transformation (Eqns.(2.41),(2.42)) which, apparently, has unphysical properties.
For example, there is no limiting velocity and the sum of two finite velocities may yield
 an infinite result. There are then two possible approaches to select the solution
corresponding to the Lorentz Transformation:
\begin{itemize}
\item[(i)] Reject the undesirable solution on the grounds that it is `unphysical'. This is
denoted by SLV2 in Table I.
\item[(ii)] Introduce another postulate that has the effect of rejecting the `unphysical'
solution.
\end{itemize}
This is the case for `Relativistic Mass Increase' in Refs.[30,31] and the `Causality Postulate'
in Ref.[25].
In fact the `unphysical' solution is not a single valued function of its arguments, so it is
rejected, in the present paper, by the Uniqueness Postulate.
\par The recent derivation of Sen~\cite{x33}, based on an ingenious gedankenexperiment involving
three parallel-moving inertial frames is remarkable in that it uses neither strong postulates
nor the Group Property. However both the Reciprocity and SLV2 Postulates were used.
In fact if the derivation of the PVAR by Mermin~\cite{x19} (also based on a sophisticated
 gedankenexperiment ) and the subsequent derivation of the Lorentz Transformation from the PVAR by Singh~\cite{x20}
 are combined, the same set of postulates as used by Sen
  (the Special Relativity Principle, the Reciprocity Postulate, Space-Time Homogeneity, Spatial Isotropy,
   and the Sign of the Limiting Velocity Squared) are invoked.
  \par It is clear that the Uniqueness Postulate replaces the STH postulate of previous
  derivations, that requires the transformation equations to be linear.
   Indeed in the direct 
  derivation of the Lorentz Transformation above the UP in the form of the trilinear equation 
  (2.3) yields immediately, on applying the condition of constant relative velocity, Eqn.(2.4),
  linear transformation equations. Uniqueness does not however imply linearity of the PVAR.
  The final equation (2.70) here retains the trilinear  form of the initial anstaz Eqn.(2.1).
  Writing the relative velocities as derivatives of the form $\delta x/\delta t$ it can be seen that 
  the PVAR is in fact the ratio of two linear
  space-time transformation equations for $\delta x$ and $\delta t$.
 \par The argument used in this paper to reject the solution with $-V^2$ in Eqn.(2.35), the non
 single-valued nature of Eqn.(2.42) for particular choices of $u$ and $v$, replaces the SLV2 
 assumption of say Sen~\cite{x33}, which rejects the solution rather because a singular 
 result is obtained for the same choices of $u$ and $v$. The economy of postulates obtained in the
 derivation presented in this paper is a consequence of the fact that the UP replaces the space-time
 postulate STH and also requires rejection the $-V^2$ solution.
 \par The Uniqueness postulate is actually implemented by assuming a multilinear form of the
 equations (trilinear in Eqns.(2.1),(2.3); quadrilinear in Eqn.(2.55)~). This is however
 not even a sufficient condition that the PVAR is single valued~\cite{x17}. Indeed
 Eqn.(2.42) is trilinear but does not, in all cases, give a unique solution for $w$. 
 All that has been shown is that a single valued multilinear solution can be found 
 that is completely determined by consistency with the remaining postulates ( effectively
 RP only, equivalent, as shown in Section 2 above, to SRP and UP).
  The interesting but unaswered question that remains is whether a single-valued solution can be found
 that does {\it not} have a multilinear form, i.e. are the Lorentz Transformations and the PVAR
 of Special Relativity the {\it only} solutions consistent with the inital postulates.
 \par Inspection of Table I shows that the minimum number of postulates necessary to 
 derive the Lorentz Transformation before the present paper is four, actually the same number
 that Einstein required for his original derivation.
 The new derivation presented here requires, in general,
 only three postulates and for the limited class of space-time points lying along the common $x$, $x'$ axis
 of the frames S, S' only \underline{\it two} postulates the Special Relativity
 Postulate and the Uniqueness Postulate
 are sufficient. 
 \par  By developing the kinematical consequences of the Lorentz Transformation
 it has been shown that, for consistency with Classical Electrodynamics, light 
 must be described by particles whose mass-equivalent is much
 smaller than their energy. That this conclusion follows from the Lorentz Transformation 
 \underline{\it alone} was pointed out, though not explicitly demonstrated, in
 Ref.[25]. Finally a remark on Pauli's discussion of kinematical derivations of
 the Lorentz Transformation ~\cite{x36}. Pauli pointed out that the Lorentz Transformation
  may be derived from the Group Property, the Reciprocity
  Postulate and the Special Relativity Postulate 
  in the form of Postulate A used here, but applied only to length measurements
 ~\cite{x37}.
 He also states that `From the group theoretical assumption it is only possible to
 derive the transformation formula but not its physical content'. In fact it has
 been demonstrated above that the physical content
  of the Lorentz Transformation actually does becomes transparent
 when its kinematical consequences are developed in detail, as in Section 3 above.
 In particular, the interpretation of $V$ as the universal limiting velocity of
 any physical object is quite general, while the identification $c=V$ for light
 leads naturally to the particle description of the latter.
 \par \underline{Acknowledgements}
 \par I should like to thank V.Telegdi for his critical reading of an earlier
 version of this paper, for several suggestions that have improved the clarity
 of the presentation and for introducing me to Pauli's book on Relativity.
 I also thank the referee for his constructive criticism and encouragement.  
 

\begin{thebibliography}{99}
\bibitem{x2}
A number of other postulates were also necessary. The most important of these were:
linearity of the equations, Spatial Isotropy and the Reciprocity Postulate. These are
discussed in relation to the present and other derivations in Section 4 .
\bibitem{x8}
This clear separation of kinematics and dynamics is not present in General Relativity,
 so it is understood that the `laws of physics' discussed in this paper exclude those
 governing gravitation.
\bibitem{x11}
That is to say that S ( S') perform similar measurements of similar measuring rods or clocks
at rest in S' ( S )
\bibitem{x15}
If the STM are associated with single quantum events (say the detection of a single photon 
reflected near the end of a measuring rod ) this will clearly not be the case, since observation by
 S  excludes observation by S' and vice-versa. It
 is understood that the STM are made within suitably short
 space and time intervals on an ensemble of `equivalent' quantum events. Depending on the
 method of measurement, one or more such `equivalent quantum events' constitute a STM.
\bibitem{x16}
Perhaps other single valued functions can be defined. Polynomials of degree n are, of course,
n-valued.
\bibitem{x17}
By a special choice of the coefficients certain `pathological' multilinear functions may be defined
that are not single-valued. For example:
\[ \alpha + \beta + \gamma - k \alpha \beta \gamma = 0 \]
has the same form as Eqn(2.1). Choosing $\alpha = 1/(k\beta) $ it is seen that the equation
is verified for all values of $\gamma$, so it is not single-valued in this variable. Except
for such special cases, the requirement of multilinearity is sufficient to ensure 
Uniqueness in the sense of Postulate B.
\bibitem{x21}
Since, clearly, $\gamma(0) = 1 $, the plus sign is taken on solving Eqns.(2.20),(2.36) for
$\gamma(v)$.
\bibitem{x23}
Since $x^2+y^2+z^2$ is rotationally invariant, this is true for a Lorentz Transformation in any direction,
not only along the $x$,$x'$ axis as for the case S $\rightarrow$ S'.
\bibitem{x35} 
 The Group Property has 3 essential aspects:
 \par 1) If the operations $A$ and $B$ are members of a Group, then so must the operation C 
 consisting of the combined operations of $A$, then $B$. In symbols $BA=C$.
 \par 2) The existence of the identity operation $I$ such that $IA=A$.
 \par 3) For each member of the group $A$, the existence of an inverse $A^{-1}$ such that 
 $A^{-1}A = I$.
\bibitem{x37}
 Although not explicitly stated by Pauli, the `Sign of the Limiting Velocity Squared'
  postulate is also needed 
 to reject the `unphysical' solution with the opposite sign of $V^2$.
\end{thebibliography}
\end{document}